# Automated Test Production Systematic Literature Mapping

Gomes, J.M.[1] and Dias, L.A.V.[1]

[1]Instituto Tecnológico de Aeronáutica - ITA

January 4, 2024

## 1 Objectives

The broader goal of this research, on the one hand, is to obtain the State of the Art in *ATP (Automated Test Production)*, to find the open questions and related problems and to track the progress of researchers in the field, and on the other hand is to list and categorize the methods, techniques and tools of *ATP* that meet the needs of practitioners who produce computerized business applications for internal use in their corporations - eventually it can be extended to the needs of practitioners in companies that specialize in producing computer applications for generic use.

## 2 Literature Systematic Mapping

### 2.1 Planning

In order to obtain an overview of the research on *ATP*, an *SLM (Systematic Literature Mapping)* is conducted here so that from this study we can perform an *SLR (Systematic Literature Review)* in order to investigate it further. We apply the method proposed by Petersen *et al.* and which we present in the Figure 1 to conduct this *SLM*[1].

We sought with this study to identify the amount and types of research and its results under the topic *ATP*. As important secondary results, we also sought to identify the discussion forums on the subject.

The question **QP1** is a filter for narrowing the scope of the research. The question **QP2** aims to identify the main discussion forums where researchers related to *ATP* publish their work or meet to present advances and update their knowledge in the area. In the question **QP3** we propose to classify the search results and identify the main types of studies related to *ATP* and categorize their contributions.

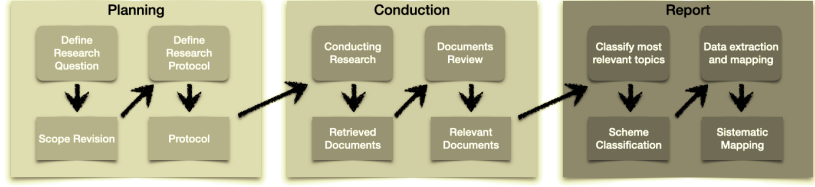

Figure 1: Steps do execute the research for an *SLM* (adapted from [1])

The search term was generated from keywords and their synonyms as presented in the Table 2. In Keele *et al.* it is pointed out that Petticrew and Roberts suggest the use of *PICOC (Population, Intervention, Comparison, Outcome, Context)* to formulate the search term in scientific publication databases [2, 3].

Two of the main considerations in designing the search protocol are that, first, we are neglecting "comparison", and are therefore using the *PIOC (Population, Intervention, Outcome, Context)* variation, because it is not part of our goal to compare different solutions to the same problem, and second, we avoid considering specific results, i.e., studies that are not aimed at the production of general-purpose computer applications, or use within an enterprise and business world context.

| # | Question |
|---|---|
| **QP1** | Is the study? |
| | *Recently published (within the last five years)*? |
| **QP2** | Which "*journals*" include studies in "*ATP*? |
| | *Or Annals of Congresses, Events, Authors, etc.* |
| **QP3** | What kinds of studies are published in *ATP*? |
| | *Categorized as listed in Figure 5* |

Table 1: Research questions for *SLM*

| **Term** | **Synonyms** | **Related to** |
|---|---|---|
| *software* | program | **Population** |
| *test* | check<br>checking<br>validation<br>verification | **Population** |
| *generation* | creation<br>inception<br>production | **Intervention** |
| *method* | methodology<br>model<br>process<br>standard | **Outcome** |
| *tool* | environment<br>framework<br>software<br>suite | **Outcome** |

Table 2: Search terms for the *SLM*

The **Context** is an extended view of the population, where we say whether the study is conducted in Academy or Industry, in which Industry segment [4]. In our case we were indifferent with regard to this aspect.

Finally we applied the search criteria of the Section 2.1 to the scientific publication databases listed in the Table 3.

```
("software" OR "program" OR "test" OR "check" OR "checking" OR "
    ↪ validation" OR "verification") AND ("generation" OR "
    ↪ creation" OR "inception" OR "production") AND ("method" OR
    ↪  "methodology" OR "model" OR "process" OR "standard" OR "
    ↪ tool" OR "environment" OR "framework" OR "software" OR "
    ↪ suite")
```
Listing 1: Search criteria for the *SLM*

## 2.2 Conduction

The scientific publication sources, listed in Table 3, are, according to Brereton *et al.*, the most relevant for Software Engineering [5].

| Name |
| --- |
| *IEEE Xplore* |
| *ACM Digital Library* |
| *Google Scholar* |
| *CiteSeerX* |
| *Inspec* |
| *ScienceDirect* |
| *EI Compendex* |
| *Springer Link* |

Table 3: Scientific publishing databases

The choice of primary research sources was based on simple and practical premises:

- use of structured search terms (using "AND", "OR", "NOT" and parentheses);
- filters to search for more recent documents; and
- filters to list relevant documents in the desired area of expertise.

Based on these criteria, the **ScienceDirect** and **ISI Web of Science** publication bases were discarded for not presenting satisfactory results for this research [1] and both **Google Scholar** and **CiteSeerX** by not providing additional filters by area or subarea of knowledge and finally **EI Compendex** and **Inspec** also did not allow access. As an alternative we used the Periodical Portal of *CAPES (Coordenação de Aperfeiçoamento de Pessoal de Nível Superior do Brasil)* [2] with the exception that the search criteria had to be adapted because of restrictions in the platform as can be seen in the Section 2.2 (as in the other platforms, additional filters were applied - see Table 7).

---

[1]The **ScienceDirect** of **Elsevier** for example restricts the search to only 8 terms, and in one exercise returned only 11 studies. Of these only one could be read in full and was still not relevant to the present study.

```
(software OR program OR test OR check OR checking OR
    ↪ validation OR verification) AND
(generation OR creation OR inception OR production)
```

Listing 2: Search criteria for the *SLM* used at *CAPES*

Then, within each publication base, filters were applied to improve the quality of the results obtained. In general, we selected only recent documents (2015 to 2020 - up to the date of the survey: April 2020). The particular criteria for each base are listed in the Tables 4 to 7.

| Filter | Value |
| --- | --- |
| **Publication title** | IEEE Access<br>IEEE Systems Journal<br>IEEE Latin America Transactions<br>IEEE Transactions on Software Engineering |
| **Indexation terms** | learning (artificial intelligence)<br>optimisation<br>neural nets<br>cloud computing<br>genetic algorithms<br>probability<br>program testing<br>search problems<br>Internet<br>resource allocation |

Table 4: IEEE Xplore filters

| Filter | Value |
| --- | --- |
| **ACM Full-Text Collection** | All journals collection |
| **Publication Title** | Search title only |

Table 5: ACM Digital Library filters

---

[2]See http://www.periodicos.capes.gov.br.

| Filter | Value |
| --- | --- |
| **Content Type** | Article |
| | Chapter |
| | Conference Paper |
| **Discipline** | Computer Science |
| **Subdiscipline** | Computer Science, general |
| | Computer Systems Organizations and Communications Networks |
| | Data Structures and Information Theory |
| | Information Systems and Communication Service |
| | Software Engineering/Programming and Operating Systems |

Table 6: Springer Link filters

| Filter | Value |
| --- | --- |
| **Type of source** | Studies |
| **Language** | English |
| **Refinement** | Pair revised journals |
| **Topic** | Computer Science |

Table 7: "Periódicos da CAPES" filters

After running the search using the Section 2.1 criteria and applying the filters listed in the Tables 4 to 7 we obtained the results listed in the Table 8 (in number of documents).

| Name | Qty. | % |
| --- | --- | --- |
| **IEEE Xplore** | 715 | 33.10 |
| **ACM Digital Library** | 308 | 14.26 |
| **Springer Link** | 709 | 32.82 |
| **Periódicos da CAPES** | 428 | 19.81 |
| **TOTAL** | 2160 | |

Table 8: Total documents retrieved from each publishing databases for the *SLM*

The selection of documents was based on Inclusion and Exclusion criteria defined iteratively during the reading of the documents found and served to determine the suitability of each one to the objectives of this work. The Inclusion Criteria are those presented in the Table 9, and in the Table 10 we have the Exclusion Criteria.

| # | Description |
|---|---|
| **CI1** | Document types: books (book excerpts), technical reports; |
| **CI2** | If several have reported the same study, only the most recent one will be considered; and |
| **CI3** | From the abstract the researcher can deduce that the article is about *ATP*. |

Table 9: Inclusion criteria for the *SLM*

| # | Description |
|---|---|
| **CE1** | The article strays from the main topic of this study which deals with *ATP* for general applications; |
| **CE2** | The topic *ATP* is not part of the article's contribution or the topic is only mentioned; and |
| **CE3** | No empirical findings or current available literature are reported. |

Table 10: Exclusion criteria for the *SLM*

Given the large number of studies (see Table 8) found we organized our work into iterative steps:

1. Reading the summary and conclusion; and

2. Selection and classification by reading the entire document.

A pre-selection was based only on the title of the document found because, as already noted by Keele *et al.*, searches of electronic databases bring a very large number of irrelevant results. During the review, other studies were rejected as being outside the scope of this study [2].

| Name | Qty. | % |
|---|---|---|
| **IEEE Xplore** | 33 | 14.73 |
| **ACM Digital Library** | 66 | 29.46 |
| **Springer Link** | 11 | 4.91 |
| **Periódicos da CAPES** | 114 | 50.89 |
| **Sub-total** | 224 | |
| *Duplicate studies* | 7 | 4.24 |
| *Rejected studies* | 52 | 31.52 |
| **TOTAL** | 165 | |

Table 11: Primary studies selction result for the *SLM*

| #      | Description                                        |
|--------|----------------------------------------------------|
| **CE1.1** | Applied to hardware;                            |
| **CE1.2** | Applied to embedded software;                   |
| **CE1.3** | Language-specific;                              |
| **CE1.4** | Does not deal with tests for general applications; |
| **CE1.5** | Not intended for general applications;          |
| **CE2.1** | Does not deal with test generation;             |
| **CE3.1** | No contribution to this study;                  |
| **CE3.2** | This is not scientific research; and            |
| **CE3.3** | Survey with old data.                           |

Table 12: Refining exclusion criteria for the *SLM*

The studies were rejected in the second selection based on a refinement of the exclusion criteria listed in the Table 10 and that we list in the Table 12. Our motivation behind these criteria, as explained in the "Objectives", is to find solutions that meet the needs of professionals who produce computerized business applications for internal use in their corporations - eventually extending to the needs of professionals in companies specializing in the production of generic computer applications.

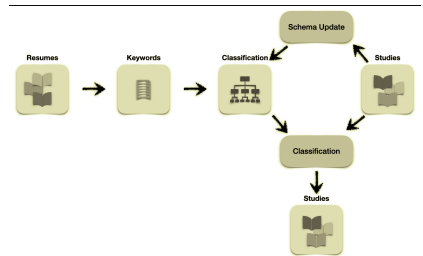

Figure 2: Classification scheme (adapted from [1])

For documents classification we adopted the Petersen *et al.* as can be seen in the Figure 2, and in the same way we adopted facets analyzing the abstracts of the studies found. We started by analyzing two main facets to classify the documents. The research type facet was based on the classification proposed by Wieringa *et al.* and summarized in the Table 13. The type of contribution is based on the interpretation of the abstracts and listed in the Table 14 [1, 6].

| Category | Description |
| --- | --- |
| Validation | The techniques investigated are new and have not been implemented in practice. Techniques used are from experimental examples and done in the laboratory. |
| Avaliation | The technique was implemented in practice and the evaluation was conducted. Demonstrated how the technique was implemented in practice and what the consequences were in terms of benefits and disadvantages. Also includes identification of problems in the industry. |
| Proposal | A solution to a problem has been proposed, either new or a significant extension of an existing technique. The potential benefits and applicability of the solution is shown by a short example or a good line of argument. |
| Philosophical | These studies outlined a new approach to pre-existing knowledge and structured the field in the form of a taxonomy or conceptual work. |
| Opinion | These studies express someone's personal opinion about a particular technique, whether it is good or not, or how things are done. They are not supported by related work or research methods. |
| Practice | They explain in what and how something was done in practice. It has to be the author's personal experience. |

Table 13: Researches types

| Category | Description |
| --- | --- |
| Metric | A system or standard of measures or measurements taken using an existing standard. |
| Tool | A device or implementation, used to perform a certain function. |
| Model | A comprehensive and systematic approach that includes theoretical principles, benefits and drawbacks, objectives, methodological guidelines and specifications, and the characteristic use of certain sets of methods and techniques. |
| Method | No particular theoretical orientation is inferred in a method. Researchers impose their own particular theoretical beliefs on an experiment when they design and implement it by applying one or more techniques. |
| Technique | A single operation or interaction in which a researcher uses one or more procedures to elicit an immediate reaction from the object of study or to shape the experiment and obtain results. |
| Procedure | An organized sequence of operations and interactions that a researcher uses to conduct an experiment. |
| Intervention | Purposefully interferes with or mitigates various aspects of the object of study and that affect the outcome by applying procedure, technique. The elements acting on the industry during a particular intervention are most often computer applications, the researcher, or both. |
| Approach | A broad way of addressing an industry concern or problem. A specific methods is not implied, but a specific set of techniques will likely come into play when trying to intervene in the industry and the problem that is the subject of the research. The procedures to be used will be determined by the delimitations of the methodological variant in which we design the study. |
| Strategy | An action plan designed to achieve an overall goal. |

Table 14: Contributions types

## 2.3 Analysis of Results

Based on the criteria listed in the Table 10 and expanded in the Table 12 we selected 165 and rejected 52 documents categorized according to the Table 15.

| Motive | Qty. |
| --- | --- |
| *Applied to hardware* | 9 |
| *Applied to embedded software* | 2 |
| *Language-specific* | 2 |
| *No contribution to this study* | 5 |
| *This is not scientific research* | 1 |
| *Does not deal with test generation* | 22 |
| *Does not deal with tests for general applications* | 3 |
| *Not intended for general applications* | 7 |
| *Survey with old data* | 1 |
| **TOTAL** | 52 |

Table 15: Rejection motives

The research questions listed in Table 1 were applied to the selected studies and we obtained the results that we list below.

**QP1 - Is the study current?**
*Recently published (within the last five years)?*

| Tipo | criteria |
|---|---|
| *2015* | 34 |
| *2016* | 32 |
| *2017* | 20 |
| *2018* | 26 |
| *2019* | 38 |
| *2020* | 15 |
| **TOTAL** | 165 |

Table 16: Publications / Year

| Tipo | Qty. | % |
|---|---|---|
| *Books* | 3 | 1.82 |
| *Conferences* | 51 | 30.91 |
| *Journals* | 111 | 67.27 |
| **TOTAL** | 165 | |

Table 17: Studies / Channel

We can observe a relative constancy of studies published on the topic in recent years, indicating that the topic is of interest and there is potential progress to be explored.

**QP2 - Which "*journals*" include studies in "*ATP*"?**
*Or Annals of Congresses, Events, Authors, etc.*

We classified how the studies were published (see Table 17) and then sought to identify where they were published to get an idea of the best *Journals* and Events where to look for information on the topic. Unfortunately, as can be seen in the Table 18 there is no specific event for this topic. In the Figure 3 and Table 19 we obtained a more satisfactory result in identifying the most relevant publications for our research.

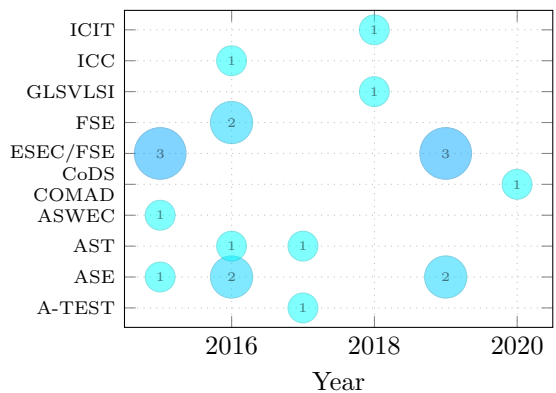

Figure 3: Participation of *ATP* Studies in Congresses

| Conference | Qty. |
| --- | --- |
| The Search-Based Software Testing (SBST) Workshop (co-located in ICSE) | 7 |
| ACM ESEC/FSE Joint European Software Engineering Conference and Symposium on the Foundations of Software Engineering | 6 |
| Internation Conference on Software Engineering | 6 |
| IEEE/ACM International Conference on Automated Software Engineering | 5 |
| ACM ICSCA Software and Computer Applications | 3 |
| ACM SIGSOFT ISSTA International Symposium on Software Testing and Analysis | 3 |
| ACM/IEEE International Conference on Automation of Software Test | 2 |
| ACM ESEC/FSE Joint European Software Engineering Conference and Symposium on the Foundations of Software Engineering | 2 |
| ACM SBES Brazilian Symposium on Software Engineering | 2 |
| Symposium on Information and Communication Technology | 2 |

Table 18: Major conferences addressing *ATP*

In the graph from Figure 3 we have a look at the interest in the topic over the course of the surveyed period in the main congresses listed.

In *Journals* we identify in the Table 19 where the topic is most frequently addressed.

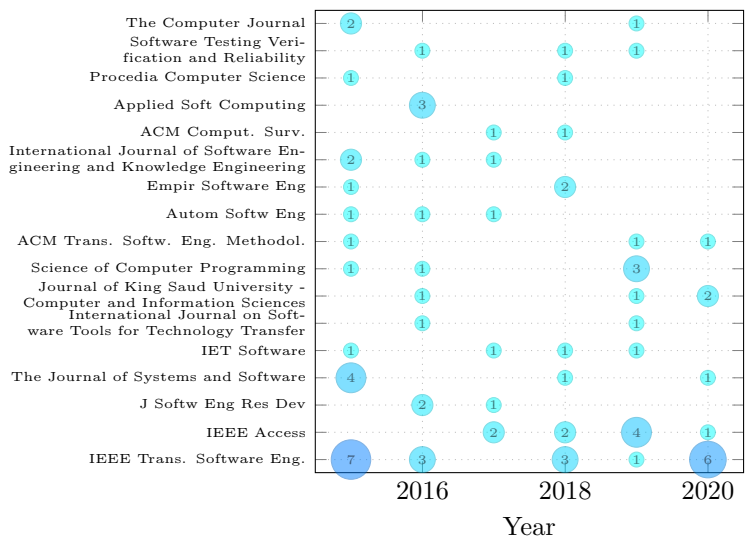

Figure 4: Participation of *ATP* Studies in Publications

| Publication | Qty. |
| --- | --- |
| *IEEE Trans. Software Eng.* | 20 |
| *IEEE Access* | 9 |
| *The Journal of Systems & Software* | 6 |
| *Science of Computer Programming* | 5 |
| *IET Software* | 4 |
| *International Journal of Software Engineering and Knowledge Engineering* | 4 |
| *Journal of King Saud University - Computer and Information Sciences* | 4 |
| *ACM Trans. Softw. Eng. Methodol.* | 3 |
| *Applied Soft Computing* | 3 |
| *Autom Softw Eng* | 3 |
| *Empir Software Eng* | 3 |
| *J Softw Eng Res Dev* | 3 |
| *Software Testing, Verification and Reliability* | 3 |
| *The Computer Journal* | 3 |

Table 19: Major Publications Addressing *ATP*

Once again we present in a graph at Figure 3 a view of the interest in the topic over the surveyed period in the main *Journals* listed.

We also sought to identify the main authors on the subject, regardless of the medium of publication. In the Table 20 we list those who presented the highest production within this research and their **H-Index** [7].

| Author | Institution | Qty. | h-idx[3] |
|---|---|---|---|
| *Harman, Mark* | *University College London* | 3 | 68 |
| *Arcuri, Andrea* | *Kristiania University College* | 9 | 39 |
| *Fraser, Gordon* | *University of Passau* | 8 | 39 |
| *McMinn, Phil* | *University of Sheffield* | 4 | 27 |
| *Zamli, Kamal Z.* | *University Malaysia Pahang* | 8 | 23 |
| *Panichella, Annibale* | *Delft University of Technology* | 3 | 27 |
| *Gargantini, Angelo* | *University of Bergamo* | 3 | 18 |
| *Vergilio, Silvia R.* | *Federal University of Paraná* | 4 | 17 |
| *Riccobene, Elvinia* | *Università di Milano* | 3 | 16 |
| *Arcaini, Paolo* | *National Institute of Informatics* | 3 | 15 |
| *Staats, Matt*[4] | *University of Luxembourg* | 3 | 15 |
| *Gay, Gregory* | *Chalmers, University of Gothenburg* | 6 | 12 |
| *Rojas, José Miguel* | *University of Leicester* | 3 | 13 |

Table 20: Main Authors in *ATP* of this Study

The Table 20 seeks to order the authors by weighting their *H-Index* and the number of publications found within the search.

**QP3 - What kinds of studies are published in *ATP*?**

*Categorized as listed in Figure 5*

We first build on the classification proposed by Wieringa *et al.* and quantify the Facet of Study Types in Figure 5 [6]. This categorization will be useful in performing a *SLR* as we qualify the studies with the desired bias for this research.

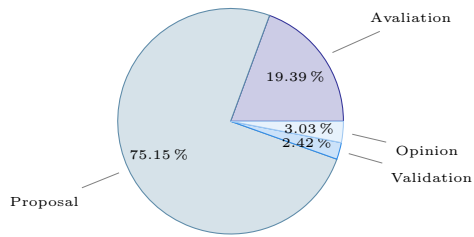

Figure 5: Types of Artifacts Generated

---
[3] Data obtained via **Google Scholar** and calculated since 2015.
[4] See **Scopus** https://www.scopus.com/results/authorNamesList.uri?sort=count-f&src=al&affilName=University+of+Luxembourg&s=AUTHLASTNAME%28Staats%29+AND+AUTHFIRST%28Matt%29+AND+AFFIL%28University+of+Luxembourg%29&st1=Staats&st2=Matt.

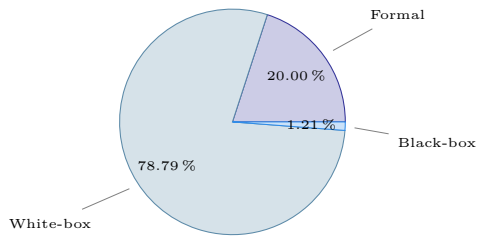

Figure 6: Types of Test Generators

In Figure 6 we categorize studies also with respect to the level of access to the source code of the program object of testing in the Verification and Validation language, where *white-box*[5] tests are applied in verification (i.e., was the correct program built?) and *black-box*[6] tests are applied in validation (i.e., was the program built correctly?). For the purposes of this research, we consider any access to the source code to mean *white-box* testing and neglect so-called *gray-box*[7] testing. We classify as *white-box* or *black-box* tests those tests that apply verification techniques through observation or direct experience.

We consider as *Formal* tests verifiable by theoretical means or pure logic, whose specifications may include expressions in various logical forms, used to write *pre* and *post* conditions, axioms of data types, constraints, temporal properties. They can represent definitions of process states, and there is a formal deduction system, enabling proofs, or other verifications (such as model checking), or both. Thus, formal specifications can be analyzed to guide the identification of appropriate test cases. According to Gaudel, these are *black-box* type tests, where the internal organization of the program under test is ignored and the strategy is based on a description of the desired properties and program behavior, grouped here by those that meet these [8] characteristics.

Also following the methodology proposed by Wieringa *et al.*, we quantified the documents by the Facet of the Artifact Type generated by the study, taking as a basis the approach presented by each and which we list in the Figure 7 [6]. In a future *SLR* we can apply qualitative aspects that will determine whether for the purposes of this research specialized (generate only the code or the test data) or general (generate both the code and the test data) approaches are the most relevant.

---

[5] Validating non-functional, internal aspects of a computer application.
[6] Validating functional and external aspects of a computer application.
[7] The combination of *white-box* and *black-box* testing methods.

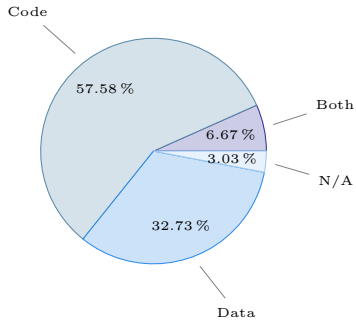

Figure 7: Types of Artifacts Generated

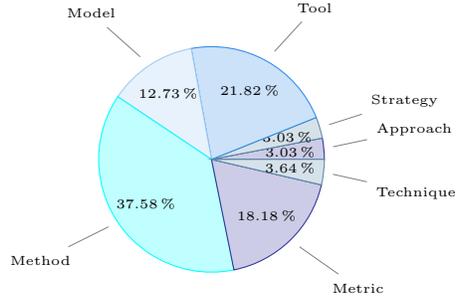

Figure 8: Contributions of the Studies

The quantification of the studies by the Contribution Type Facet (see Figure 7) was important for the qualification and selection of the most relevant studies to meet the objectives of the present research and the quantified can be observed in the Figure 8.

## 3 Results

Was presented the elementary results of a *SLM* applied to finding relevant studies in *ATP*. This review applied the methodology of Petersen *et al.* with elements of Keele *et al.*, Brereton *et al.* [1, 2, 5]

### 3.1 Conclusions

Based on the research questions developed in the Table 1 the conclusions is:

#### 3.1.1 Is the study current?

Was ensured that the studies were recent by restricting our search to the last 5 years and we can observe in the Table 16 an even distribution of studies across the surveyed period.

### 3.1.2 Which "*journals*" include studies in ATP?

In the Table 17 it can be seen the large concentration of studies published in conferences and "*Journals*" and this led us to list the main conferences (see Table 18) and the main publications (see Table 19). The relatively small number of books on the subject, in our view, is due to the innovative characteristics under which the fields of engineering and computer science live today.

### 3.1.3 What categories of studies are published in ATP?

Of particular interest to our research on *ATP*, the types of studies that stood out the most can be seen in Figure 8. The concentration in practical aspects, as tools, methods, models and metrics leads us to conclude that the maturity the subject is now in Academy. The generation of both code and data is addressed by the studies, and this is a guarantee that we're covering all aspects of the subject.

## 3.2 Future Work

This work aims to prepare ground for a *SLR* where it will determine the challenges in applying generative testing techniques and evaluate the solutions intended to be applied in future work.

# Acronyms

**ATP** Automated Test Production - pages: 1, 2, 6, 10, 13, 15, 16

**CAPES** Coordenação de Aperfeiçoamento de Pessoal de Nível Superior do Brasil - page: 4

**PICOC** Population, Intervention, Comparison, Outcome, Context - page: 2

**PIOC** Population, Intervention, Outcome, Context - page: 2

**SLM** Systematic Literature Mapping - pages: 1, 15

**SLR** Systematic Literature Review - pages: 1, 13, 14, 16